# Graphs, Ideal Flow, and the Transportation Network

Kardi Teknomo
Ateneo de Manila University
kteknomo@ateneo.edu

## Introduction

In this lecture, I discuss the mathematical relationship between network structure and network utilization of transportation network. Network structure means the graph itself. Network utilization represent the aggregation of trajectories of agents in using the network graph. I show the similarity and relationship between the structural pattern of the network and network utilization.

An ideal flow is an aggregation of the trajectories of results of random walk with uniformly distributed flow over space and time on a network graph, which maximize the entropy.

$$\max_{p_j} H = -\sum_{j=1}^{k} p_j \log_2 p_j \quad s.t. \quad \sum_{j=1}^{k} p_j = 1$$

I found out that the ideal flow is invariant from the number of agents and the length of simulation. This implies that ideal flow matrix (which is clearly a network utilization) depends only on the network structure. We also how that ideal flow matrix is always premagic matrix. Premagic matrix characterizes flow conservation on nodes.

## Definitions

The structure of a network is modelled by a graph *G=(V,E)* and can be defined by several matrices.

- Adjacency matrix $\mathbf{A} = [\mathbf{a}_{ij}]$ where $\mathbf{a}_{ij} = \begin{cases} 1 & if\ (i,j) \in E \\ 0 & otherwise \end{cases}$

- Path matrix $\mathbf{P} = [\mathbf{p}_{st}]$ where $\mathbf{p}_{st} = \begin{cases} \delta(s,t) & \text{if there is a path from } s \text{ to } t \\ +\infty & \text{otherwise} \end{cases}$

- External matrix $\mathbf{E} := \mathbf{P} - \mathbf{A}$

The utilization of a network is modeled based on the agent's trajectories. A *trajectory* is a path of a moving agent (such as a vehicle, pedestrian, product, or a simple particle) within a specified observation time period from $t_1$ to $t_2$, where $t_1 < t_2$. The path is often collected at discrete points using tracking devices. Given a set of trajectories, it is possible to analyse the trajectories in three levels:
1. Matrix-set level
2. Matrix-count level
3. Matrix-structure level

The basic structure come from the matrix-set level, where each element represents a set of trajectories, distinguished by their identification numbers. Matrix-count level is derived from matrix-set level by counting the number of trajectories in each matrix-set element. Most of the traditional matrix analyses are Flow matrix and Origin Destination Matrix, which are on the matrix-count level. Matrix-structure level deals with binarized matrices that are derived from the matrix-count. We can define several matrices to represent network utilization.

- Flow-set $\tilde{\mathbf{f}}_{ij}$ (an element of the flow matrix $\tilde{\mathbf{F}}$) as a set of trajectory identification numbers that pass through the direct link $\vec{ij}$.

$$\tilde{\mathbf{f}}_{ij} := \left\{ tr \in Tr \mid tr = \left(v_1, v_2, ..., v_h = i, v_{h+1} = j, ..., v_k\right) \exists h,\ 1 \leq h < k \right\}$$

Where $Tr$ is the set of trajectories.

- The generalized OD matrix-set $\tilde{\mathbf{D}}$ is the matrix whose element $\tilde{\mathbf{d}}_{st}$ gives the set of trajectories that pass through node *s* to node *t* either directly (through a single link) or indirectly (through several links).

$$\tilde{\mathbf{d}}_{st} := \left\{ tr \in Tr \mid tr = \left(v_1, v_2, ..., v_p = s, ..., v_q = t, ..., v_k\right) \exists p, q,\ 1 \leq p < q \leq k \right\}.$$

- An indirect flow between a given OD pair *s-t* represents the amount of flow from *s* to *t* passing through more than 1 link. We formalize this concept and use the matrix-set $\tilde{\mathbf{L}}$ to describe the trajectory sets of indirect flow:

$$\tilde{\mathbf{I}}_{st} := \left\{ tr \in Tr \mid tr = \left(v_1, v_2, ..., v_p = s, ..., v_r, ..., v_q = t, ..., v_k\right) \exists p, q, r,\ 1 \leq p < r < q \leq k \right\}.$$

- The indirect flow can be partitioned further to distinguish between cases where a direct link is available (*alternative route flow* which is represented by the matrix $\mathbf{T}$) and where there is no direct link from a given node *s* to a given node *t* (*substitute route flow* which is represented by $\mathbf{T^C}$). The mutual exclusivity of these two gives rise to the following equations.

$$\tilde{\mathbf{L}} := \tilde{\mathbf{T}} + \tilde{\mathbf{T}}^{\mathbf{C}}$$

*Premagic matrix* is a square matrix where the vector row sum is equal to the transpose of vector column sum

$$\left(\mathbf{Fj}\right)^T = \mathbf{j}^T \mathbf{F}$$

An *ideal flow* is the infinite limit of relative aggregated count of random walk agents' trajectories on a network graph distributed over space and time. *Generalized Ideal flow* matrix $\mathbf{F}$ is defined by 3 properties:

1. $\mathbf{F}$ is non-negative matrix.
2. The main diagonal entries of $\mathbf{F}$ are zero.
3. $\mathbf{F}$ is pre-magic matrix.

*Standard Ideal Flow* or Uniform distribution Ideal flow matrix $\mathbf{F}$ is defined by 4 properties. In addition to the 3 properties of the Generalized Ideal flow, we have

4. $\mathbf{F}$ satisfies equal outflow constraint.

## Results

Given a generalized origin destination matrix $\mathbf{D} = [\mathbf{d}_{st}]$, a flow matrix $\mathbf{F} = [\mathbf{f}_{ij}]$ and the network graph represented by its adjacency matrix $\mathbf{A} = [\mathbf{a}_{ij}]$, the following relationships hold:

$$\mathbf{F} \leq \mathbf{A} \circ \mathbf{D} \leq \mathbf{D}$$

**Theorem-1**: Given a network structure's adjacency matrix $\mathbf{A}$, binarized distance matrix $\hat{\mathbf{P}}$ and binarized external matrix $\hat{\mathbf{E}}$, and given the network utilization quantified by a flow matrix $\mathbf{F}$, OD matrix $\mathbf{D}$, indirect flow matrix $\mathbf{L}$ and alternative route flow matrix $\mathbf{T}$, the following relationships are valid and are equivalent:

$$\mathbf{L} = \mathbf{T} + \hat{\mathbf{E}} \circ \mathbf{D}$$
$$\mathbf{A} \circ \mathbf{D} = \mathbf{D} - \hat{\mathbf{E}} \circ \mathbf{D}$$
$$\mathbf{D} = \hat{\mathbf{P}} \circ \mathbf{D}$$
$$\mathbf{F} = \mathbf{A} \circ \mathbf{D} - \mathbf{T}$$

We also performing random walk of N agents on the graph. On each node, the agent takes a random choice to the available directed edge. The agents keep moving until time T➔∞ is achieved. We record trajectory data of the agents.

**Theorem 2**: When we increase the product of the number of agents and the simulation time to infinite term, the flow will change but the relative flow would be asymptotically converge to a certain value, which we called as *ideal flow*.

$$\mathbf{F} = \lim_{N \cdot T \to \infty} \frac{\mathbf{R}}{\min \mathbf{R}} \; if \; \mathbf{r}_{ij} \neq 0$$

When N*T is large to fill the irreducible aperiodic network we will have asymptotic values of the ideal flow. Thus, ideal flow is an aggregation of random walk on network (i.e. network utilization) but the values depends only on network structure.